# Stopping Light All-Optically

Mehmet Fatih Yanik and Shanhui Fan

Ginzton Laboratory, Stanford University, Stanford, California 94305

We show that light pulses can be stopped and stored all-optically, with a process that involves an adiabatic and reversible pulse bandwidth compression occurring entirely in the optical domain. Such a process overcomes the fundamental bandwidth-delay constraint in optics, and can generate arbitrarily small group velocities for light pulses with a given bandwidth, without the use of any coherent or resonant light-matter interactions. We exhibit this process in optical resonator systems, where the pulse bandwidth compression is accomplished only by small refractive index modulations performed at moderate speeds.

The ability to drastically slow down the propagation speed of light, and to coherently stop and store optical pulses, holds the key to the ultimate control of light, and has profound implications for optical communications [1] and quantum information processing [2,3]. In order to reduce the group velocity of light coherently, there are two major approaches, which employ either electronic or optical resonances. Using electronic resonances in atomic systems, the group velocity of light can be decreased by several orders of magnitude [4]. Furthermore, with the use of quantum interference schemes such as the Electromagnetically Induced Transparency (EIT), the absorption at some electronic resonances can be strongly suppressed [5]. Dramatic slow down or even complete stop of light pulses can then be accomplished by converting the optical signal into coherent electronic states [6-13]. The use of electronic states to coherently store the optical information, however, imposes severe constraints on the operating conditions. As a result, only a few very special and delicate electronic resonances available in nature possess all the required properties. All the demonstrated operating bandwidths are far too small to be useful for most purposes. The wavelength ranges where such effects can be observed are also very limited. Furthermore, while promising steps have been taken for room temperature operation in solid-state systems, it still remains a great challenge to implement such schemes on-chip with integrated optoelectronic technologies [12-13].

Consequently, it is of great interest to pursue the control of light speed using optical resonances in photonic structures including dielectric micro-cavities [14] and photonic crystals [15-17]. Photonic structures can be defined by lithography and designed to operate at any wavelength range of interest. Ultra-high quality-factor cavities have been realized on semiconductor chips [18], and group velocities as low as $10^{-2}c$ for pulse propagation with negligible distortion have been experimentally observed in photonic crystal waveguide band edges [19] or with Coupled Resonator Optical Waveguides (CROW) [20-22]. Nevertheless, such structures are fundamentally limited by the so-called delay-bandwidth product [23] – the group delay from an optical resonance is inversely proportional to the bandwidth within which the delay occurs. Therefore, for a given optical pulse with a certain temporal duration and corresponding frequency bandwidth, the minimum group velocity achievable is limited. In a CROW waveguide structure, for example, the minimum group velocity that can be accomplished for pulses at $10 Gbit/s$ rate with a wavelength of $1.55 \mu m$ is no smaller than $10^{-2}c$. For this reason, up to now, photonic structures could not be used to stop light.

Here we introduce a set of general criteria to overcome the fundamental limit imposed by the delay-bandwidth product in optics. These criteria enable one to generate arbitrarily small group velocities for optical pulses with a given bandwidth, while preserving all the coherent information entirely in the optical domain. We show that these criteria can be achieved in optical resonator systems using only small refractive-index modulations performed at moderate speeds, even in the presence of losses. In addition, since the bandwidth constraints occur in almost all physical systems that use resonance enhancement effects, our approach to overcome such constraints is applicable to a wide range of systems and applications.

In order to coherently stop an optical pulse with a given bandwidth in an all-optical system, the following criteria must be satisfied:

(a) The system must possess large tunability in its group velocity. To allow for an optical pulse with a given bandwidth to enter the system, the system must possess an initial state with a sufficiently large bandwidth (i.e. a large group velocity as required by the delay-bandwidth product) in order to accommodate all the spectral components of the pulse. We design a system such that a small refractive-index shift can change the group velocity by many orders of magnitude, and that the group velocity reduction is independent of losses.

(b) The tuning of the system needs to be performed in a manner such that the bandwidth of the pulse is reversibly compressed. Such bandwidth compression is necessary in order to accommodate the pulse as the system bandwidth is reduced. Thus, the tuning process must occur while the pulse is completely in the system, and must be performed in an adiabatic [24] fashion to preserve all the coherent information encoded in the original pulse. In our design, we use a translationally invariant refractive-index modulation to conserve the wavevector information. The modulation accomplishes a coherent frequency conversion process for all spectral components, and reversibly compresses the bandwidth of the incident pulse.

We exhibit these concepts in the system shown in Figure 1, which consists of a periodic array of coupled cavities. Each unit cell of the periodic array contains a waveguide-cavity A, which is coupled to the nearest neighbor unit cells to form a coupled resonator optical waveguide, and one or more side-cavities $B_1$ and $B_2$, which couples only to the cavities in the same unit cell. The side-cavities in adjacent unit cells are placed in an alternating geometry in order to prevent coupling between them.

For the simple case where only a single side-cavity B exists in each unit cell, the dynamics of the field amplitudes $a_n$, $b_n$ for cavities A and B in the n$^{th}$ unit cell can be expressed using coupled mode theory, as:

$$\frac{da_n}{dt} = i\omega_A a_n + i\alpha(a_{n-1} + a_{n+1}) + i\beta b_n - \gamma_A a_n \qquad (1)$$

$$\frac{db_n}{dt} = i\omega_B b_n + i\beta a_n - \gamma_B b_n \qquad (2)$$

Here $\alpha$, $\beta$ are the coupling constants between the pairs of cavities A-A and A-B respectively. $\omega_A$ and $\omega_B$ are the resonance frequencies, and $\gamma_A$ and $\gamma_B$ are the loss rates for the cavities A and B respectively.

Since the system has translational symmetry along the waveguide, the frequencies $\omega_{\pm,k}$ for the eigenstates of the system can be related to a wavevector $k$ as

$$\omega_{\pm,k} = \frac{1}{2}\left[\omega_{A,k} + \omega_B + i(\gamma_A + \gamma_B) \pm \sqrt{(\omega_{A,k} - \omega_B + i(\gamma_A - \gamma_B))^2 + 4\beta^2}\right] \qquad (3)$$

where $\omega_{A,k} = \omega_A + 2\alpha\cos(k\ell)$ represents the frequency band of the waveguide by itself. For concreteness, we focus on the lower band $\omega_{-,k}$, which has a group velocity at the band center of

$$v_{g-} = \text{Re}\left[\frac{d\omega_{-,k}}{dk}\right]_{k=\pi/2\ell} = \alpha\ell\,\text{Re}\left[1 - \frac{\Delta + i(\gamma_A - \gamma_B)}{\sqrt{(\Delta + i(\gamma_A - \gamma_B))^2 + 4\beta^2}}\right] \qquad (4)$$

with $\Delta \equiv \omega_A - \omega_B$. When $\Delta \ll -|\beta|$, the lower band exhibits a large group velocity ($v_g \simeq 2\alpha\ell$) and a large bandwidth (Figure 2a). When $\Delta \gg |\beta|$ (Figure 2b), the group velocity at the band center (and also the bandwidth) is reduced by a ratio of $\beta^2/\Delta^2 + (\gamma_A - \gamma_B)^3/(4\Delta^3)$. Importantly, the group velocity becomes independent of loss when $\gamma_A$ and $\gamma_B$ are equal. Also, by increasing the number of side-cavities in each unit cell as shown in Figure 1, the minimum achievable group velocity at the band center can be further reduced to $2\alpha\ell \prod_{i=1}^{r} (\beta_i/\Delta)^2$ where $r$ is the number of the side-cavities in each unit cell, and $\beta_i$ is the coupling constant between the $(i\text{-}1)^{th}$ and $i^{th}$ side-cavities as shown in Figure 1. Thus, the group velocity can be reduced exponentially with linear increase in system complexity, and significant group velocity tuning can be accomplished with the use of small refractive index variation that changes the resonant frequencies $\omega_A$ and $\omega_B$.

In this system, a pulse can be stopped by the following dynamic process: We start with $\Delta \ll -|\beta|$, such that the lower band has a large bandwidth. By placing the center of $\omega_{-,k}$ at the pulse carrier frequency $\omega_0$ (Figure 2a), the lower band can accommodate the entire pulse, with each spectral component of the pulse occupying a unique wavevector. After the pulse is completely in the system, we vary the resonance frequencies until $\Delta \gg |\beta|$ (Figure 2b), at a rate that is slow enough compared with the frequency separation between the lower and the upper bands. (The frequency separation reaches a minimum value of $2|\beta|$ when $\Delta = 0$). The modulation of the cavity resonances preserves translational symmetry of the system. Therefore, cross talk between different wavevector components of the pulse is prevented during the entire tuning process. Also, the slow modulation rate ensures that each wavevector component of the pulse follows only the lower band $\omega_{-,k}$, with negligible scattering into the upper band $\omega_{+,k}$ (i.e. the system evolves in an adiabatic [24] fashion). Consequently, the pulse bandwidth is reversibly compressed via energy exchange with the modulator, while all the information encoded in the pulse is preserved. We note that, for such frequency compression to occur, the modulation does not need to follow any particular trajectory in time except being adiabatic, and can have a far narrower spectrum than the bandwidth of the incident pulse.

We implement the system presented above in a photonic crystal structure that consists of a square lattice of dielectric rods $(n=3.5)$ with a radius of $0.2a$, ($a$ is the lattice constant) embedded in air $(n=1)$ (Figure 3). The photonic crystal possesses a band gap for TM modes with electric field parallel to the rod axis. Decreasing the radius of a rod to $0.1a$ generates a single mode cavity with resonance frequency at $\omega_0 = 0.3224 \cdot (2\pi c/a)$. Coupling between two neighboring cavities A and between the adjacent cavities A and B occur through barriers of three rods ($\ell = 4a$), with coupling constants of $\alpha = \beta = 0.00371 \cdot (2\pi c/a)$. The resonant frequencies of the cavities are tuned by refractive index modulation of the cavity rods. We simulate the entire process of stopping light for $N=100$ pairs of cavities with finite-difference-time-domain method, which solves Maxwell's equations without approximation. The waveguide is terminated by introducing a loss rate in the last cavity by an amount equal to the coupling constant $\alpha$, which provides a perfectly absorbing boundary condition for the waveguide mode. The dynamic process for stopping light is shown in Figure 3a. We generate a Gaussian pulse by exciting the first cavity (The process is in fact independent of the pulse shape one chooses). The excitation reaches its peak at $t = 0.8t_{pass}$, where $t_{pass}$ is the traversal time of the pulse through the waveguide by itself. While the pulse is generated, the waveguide is in resonance with the pulse frequency while the side-cavities are kept detuned. Thus, the field is concentrated in the waveguide region (Figure 3b, $t = 0.8t_{pass}$), and the pulse propagates inside the waveguide at a relatively high group velocity speed of $2\alpha\ell$. After the pulse is generated, we gradually tune the side-cavities into resonance with the pulse while de-tune the waveguide out of resonance. At the end of this process, the field is almost completely transferred from the waveguide to the side-cavities (Figure 3b, $t = 2.0t_{pass}$), and the group velocity becomes greatly reduced. Empirically, we have found that the use of a simple modulation ($\exp[-t^2/t_{mod}^2]$) with the rise and fall times of $t_{mod} = 10/\beta$ is sufficient to satisfy the adiabatic tuning condition. Although in principle modulation of only either the side-cavities or the waveguide-cavities is necessary, we have modulated both cavities with equal strength to minimize the frequency shift required for a given group velocity reduction. With the waveguide out of resonance, the pulse is held in the side-cavities (Figure 3b, $t = 5.0t_{pass}$), and shows almost no forward motion over the time period of $3t_{pass}$ except phase change. Then, after an arbitrarily selected delay of $5.0t_{pass}$, the pulse is released by the same index modulation process above repeated in reverse, with the side-cavities gradually detuned off resonance while the waveguide tuned into resonance (Figure 3b, $t = 6.5t_{pass}$). The pulse intensity as a function of time in the last cavity of the waveguide is plotted in Figure 3a, and shows the same temporal shape as both the pulse that propagates through the waveguide by itself, and the initial pulse as recorded in the first cavity of the waveguide. Thus, our simulation indeed demonstrates that the pulse is perfectly recovered without distortion after the intended delay of $5.0 t_{pass}$, and the FDTD simulation agrees very well with the coupled mode theory analysis. In the FDTD simulations, we choose an index modulation of 8% and a modulation rate of $5GHz$ to make the total simulation time feasible. The simulation demonstrates a group velocity of $10^{-4}c$ for a $4ps$ pulse at $1.55\mu m$ wavelength. Such a group velocity is at least two orders of magnitude smaller than the minimum group velocity achievable for such a pulse in any conventional slow-light structure.

In practical optoelectronic devices [25], the modulation strength ($\delta n/n$) is typically on the order of $10^{-4}$ at a maximum speed exceeding $10GHz$. Since such modulation strength is far weaker compared with what is used here in the FDTD simulation, the coupled mode theory should apply even more accurately in the realistic situation. Therefore, using coupled mode theory, we have simulated the structure shown in Figure 1 with two side-cavities coupled to each waveguide-cavity. We use coupling constants of $\beta_1 = 10^{-5}\omega_A$ and $\beta_2 = 10^{-6}\omega_A$, a maximum index shift of $\delta n/n = 10^{-4}$, and assume a cavity loss rate of $\gamma = 4 \cdot 10^{-7}\omega_A$ that has been measured in on-chip micro-cavity structures [18]. A waveguide-cavity coupling constant of $\alpha = 10^{-5}\omega_A$ is used to accommodate a $1 ns$ pulse. Here, the bandwidth compression process occurs in two stages, first by transferring the field from the cavities A to $B_1$, and then from the cavities $B_1$ to $B_2$. At the end of this compression process, the group velocity reduces to below *0.1 meters per second*. The same process repeated in reverse recovers the original pulse shape without any distortion in spite of the significant loss present. At such ultra-slow speeds, the pulses stay stationery in the side-cavities and experience negligible forward propagation. The storage times then become limited only by the cavity lifetimes. Importantly, the storage times are also independent of the pulse bandwidths, which enable the use of ultra-high quality-factor microcavities to store short (large bandwidth) pulses coherently, by overcoming the fundamental bandwidth constraints in ultra-high Q cavities. The performance can be further improved by the use of gain mediums in the cavities [26] to counteract the losses.

The required number of the cavities is determined by the bandwidth of the pulse, which sets the maximum speed in the waveguide, and the duration of the modulation during the first stage of the field transfer, which sets the distance that the pulse travels before its speed is reduced. Thus, by using a relatively large coupling between the side-cavities $B_1$, and waveguide-cavities A, a fast slow-down of the pulse is achieved without violating adiabaticity, which reduces the propagation distance of the pulse significantly. For the two-stage system presented above, to accomplish the entire process of slowing down and recovering, a waveguide with a total length of 120 microcavities modulated at a maximum of 1*GHz* has been sufficient. Thus chip scale implementation of such systems is foreseeable. Since the group velocity reduction is translationally invariant in space, pulse length does not change as the pulse comes to a halt. Multiple pulses can be held simultaneously along such a system, and desired pulses can then be released on demand. This capability might enable controlled entanglement of networks of quantum systems in distant microcavities via photons, thus opening up the possibility of chip scale quantum information processing with photons. The ultra-low group velocity could also be used to significantly enhance nonlinear effects over the entire bandwidths of pulses.

The work was supported in part by NSF Grant No. ECS-0200445. The simulations were performed at the Pittsburgh Supercomputing Center through the support of a NSF-NRAC grant.

**FIGURE CAPTIONS:**

**Figure 1** Schematic of a tunable micro-cavity system used to stop light. The disks represent cavities, and the arrows indicate available evanescent coupling pathways between the cavities. The system consists of a periodic array of coupled cavities. Each unit cell of the array contains a waveguide-cavity A, which couples to nearest neighbor cells via evanescent coupling with a coupling strength $\alpha$. Each waveguide-cavity A is also coupled to either one or more side-cavities (with coupling strength $\beta_i$'s). The figure shows the case with two side-cavities, labeled as $B_1$ and $B_2$.

**Figure 2** Schematic of the frequency bands $\omega_+$ and $\omega_-$ for the system shown in Figure 1 with a single side-cavity in each unit cell. $\omega_A$ and $\omega_B$ are the resonance frequencies for the waveguide-cavities A and the side-cavities B, and $\beta$ is the coupling constant between them. The widths of the lines represent the widths of the frequency bands. **(a)** $\omega_A - \omega_B \ll -|\beta|$. The frequency band $\omega_-$ exhibits a large bandwidth centered at the pulse frequency $\omega_0$. **(b)** $\omega_A - \omega_B \gg |\beta|$. The frequency band $\omega_-$ exhibits a small bandwidth.

**Figure 3** Propagation of an optical pulse through a coupled micro-cavity complex in a photonic crystal system as the resonant frequencies of the cavities are varied. The photonic crystal consists of 100 cavity pairs. The pulse is generated by exciting the first cavity, and the excitation reaches its peak at $t = 0.8 t_{pass}$, where $t_{pass}$ is the traversal time of the pulse through the waveguide by itself. Fragments of the photonic crystal are shown

in part b. The three fragments correspond to cavity pairs 3-6, 56-60, 97-100. The dots indicate the positions of the dielectric rods. The black dots represent the cavities. (**a**) The dashed green and black lines represent the variation of $\omega_A$ and $\omega_B$ as a function of time, respectively. The blue solid line is the intensity of the incident pulse as recorded in the first waveguide-cavity. The red dashed and solid lines represent the intensity in the last waveguide-cavity, in the absence and in the presence of group velocity reduction, respectively. The group velocity reduction occurs from $1.0t_{pass}$ to $1.5t_{pass}$. The pulse is then held near stationary in the system until $6.0t_{pass}$. Afterwards, the pulse is completely released at $6.5t_{pass}$. Open circles are FDTD results, and the red and blue lines are from coupled mode theory. (**b**) Snapshots of the electric field distributions in the photonic crystal at the indicated times. Red and blue represent large positive and negative electric fields, respectively. The same color scale is used for all the panels.

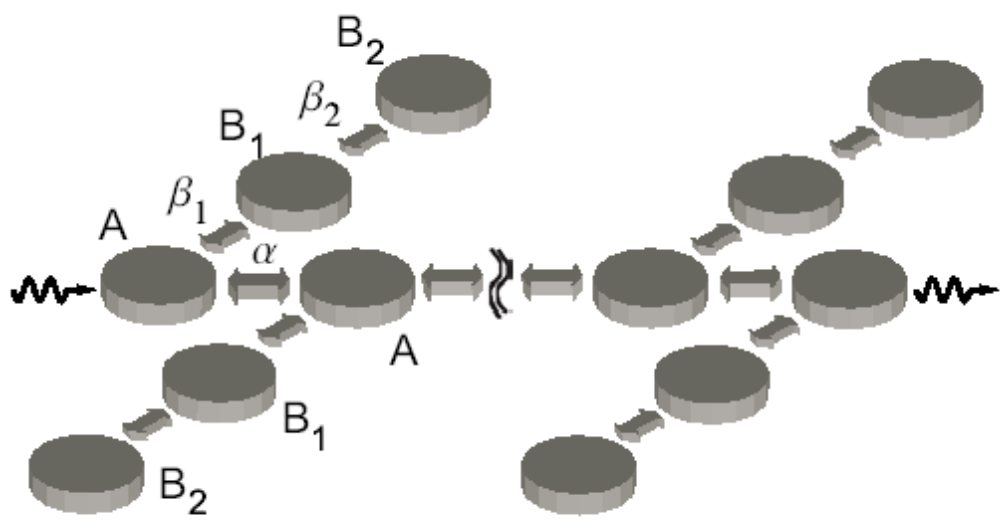

Figure 1. Mehmet F. Yanik, Shanhui Fan

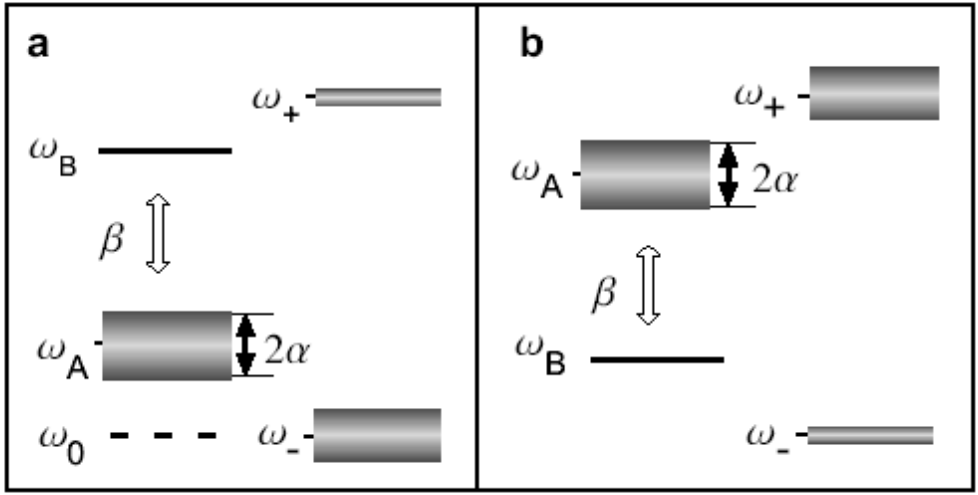

Figure 2. Mehmet F. Yanik, Shanhui Fan

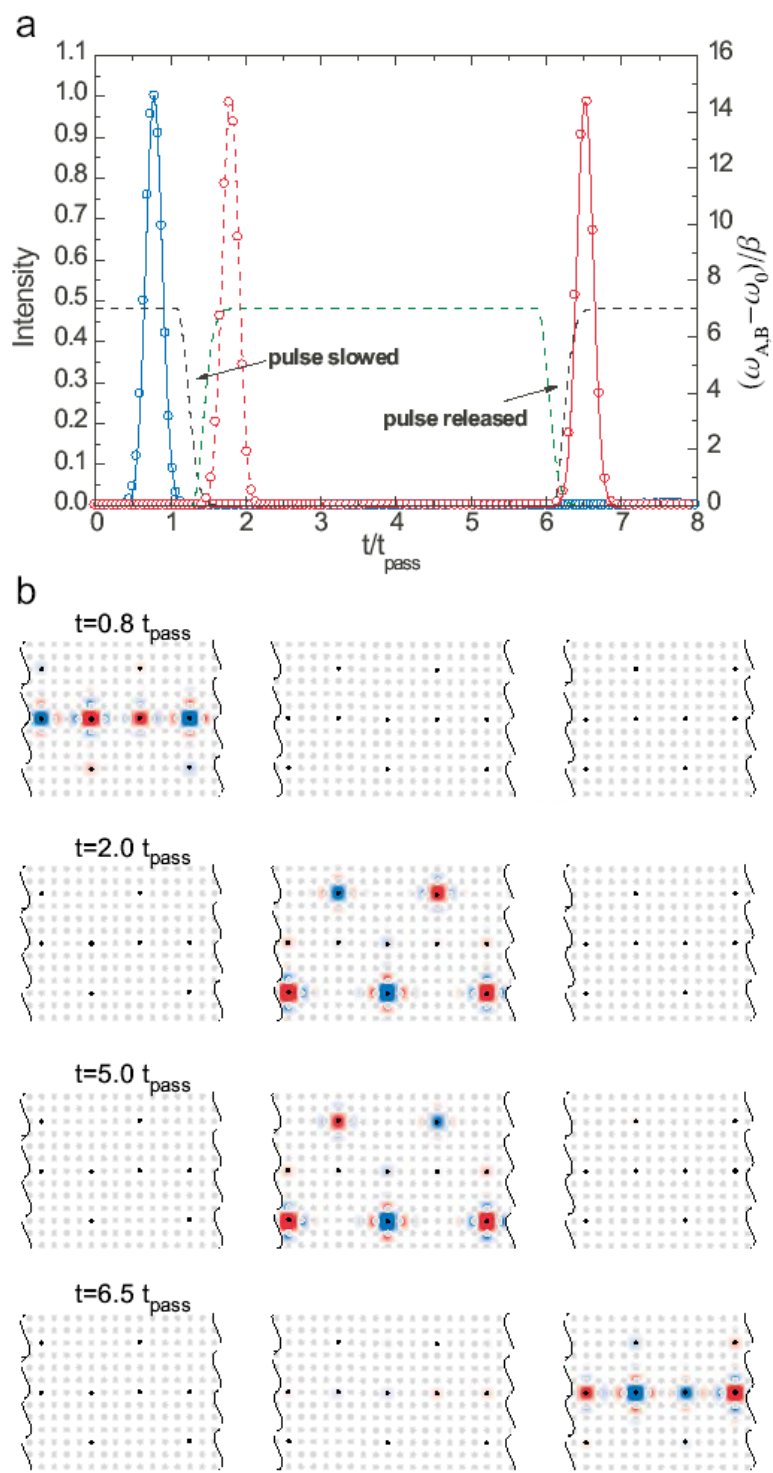

Figure 3. Mehmet F. Yanik, Shanhui Fan